\documentclass[a4paper,fleqn,usenatbib]{mnras}

\usepackage[T1]{fontenc}
\usepackage{ae,aecompl}

\usepackage{graphicx}	
\usepackage{amsmath}	
\usepackage{amssymb}	
\usepackage{epstopdf}
\usepackage{textcomp}
\usepackage{multirow}
\usepackage{media9}

\newcommand{\kms}{\,km\,s$^{-1}$}	

\title[Dynamics of A3266 -- I. An Optical View]{Dynamics of Abell 3266 -- I. An Optical View of a Complex Merging Cluster}

\author[S. Dehghan et al.]{
Siamak~Dehghan,$^{1,2}$\thanks{E-mail: siamak.dehghan@vuw.ac.nz}
Melanie~Johnston-Hollitt,$^{1,2}$
Matthew~Colless$^{3}$
\newauthor{
and Rowan~Miller$^{1}$}
\\
$^{1}$School of Chemical and Physical Sciences, Victoria University of Wellington, P.O. Box 600, Wellington 6140, New Zealand\\
$^{2}$Peripety Scientific Ltd., P.O. Box 11355, Manners Street, Wellington 6142, New Zealand\\
$^{3}$Research School of Astronomy and Astrophysics, Australian National University, Canberra, ACT 2611, Australia
}

\date{Accepted XXX. Received YYY; in original form ZZZ}

\pubyear{2016}

\begin{document}
\label{firstpage}
\pagerange{\pageref{firstpage}--\pageref{lastpage}}
\maketitle

\begin{abstract}
We present spectroscopy of 880 galaxies within a 2-degree field around the massive, merging cluster Abell 3266. This sample, which includes 704 new measurements, was combined with the existing redshifts measurements to generate a sample of over 1300 spectroscopic redshifts; the largest spectroscopic sample in the vicinity of A3266 to date. We define a cluster sub-sample of 790 redshifts which lie within a velocity range of 14,000 to 22,000~\kms and within 1 degree of the cluster centre. A detailed structural analysis finds A3266 to have a complex dynamical structure containing six groups and filaments to the north of the cluster as well as a cluster core which can be decomposed into two components split along a northeast-southwest axis, consistent with previous X-ray observations. The mean redshift of the cluster core is found to be $0.0594 \pm 0.0005$ and the core velocity dispersion is given as $1462^{+99}_{-99}$~\kms. The overall velocity dispersion and redshift of the entire cluster and related structures are $1337^{+67}_{-67}$~\kms and $0.0596 \pm 0.0002$, respectively, though the high velocity dispersion does not represent virialised motions but rather is due to relative motions of the cluster components. We posit A3266 is seen following a merger along the northeast southwest axis, however, the rich substructure in the rest of the cluster suggests that the dynamical history is more complex than just a simple merger with a range of continuous dynamical interactions taking place. It is thus likely that turbulence in A3266 is very high, even for a merging cluster.
\end{abstract}

\begin{keywords}
galaxies: clusters: general -- galaxies: clusters: individual: A3266 -- galaxies: distances and redshifts
\end{keywords}



\section{Introduction}

Matter is not uniformly distributed in the Universe, rather it is concentrated in large filaments of galaxies which are tangled together forming a vast web-like structure known as the `cosmic web' \citep{bkp96}. Within the cosmic web gigantic conglomerations of galaxies known as superclusters reside. Superclusters contain numerous galaxy clusters and groups, and thousands of individual galaxies. Studying large-scale structures such as superclusters and their building blocks, galaxy clusters, provides clues to understanding the initial conditions and fluctuations of the early Universe, and the way it has evolved over time.

The Horologium-Reticulum supercluster (HRS) is a super massive ($10^{17}$ M$_{\odot}$) and non-gravitationally bound structure with a mean redshift of about 0.06 \citep{frc05,frc06}. The supercluster has an angular span of $12\degr \times 12\degr$, and includes about 5000 groups and 34 clusters. One HRS member, Abell 3266 is one of the largest galaxy clusters in the Southern Hemisphere. A3266 has been well studied over the past two decades and a wealth of multi-wavelength data, from X-ray \citep{dm99,fhm06}, to optical \citep{csc01}, and radio \citep{rr90,bvc16}, are accessible.

Abell 3266 is known to be a merging system; examination of its soft-band X-ray image reveals asymmetrical and elongated emission from the Intra-Cluster Medium \citetext{ICM, \citealp{fhm06}}, typical for merging and highly disrupted systems. However, detailed structure analysis of the cluster has been somewhat hindered due to the lack of sufficient spectroscopic redshifts. Consequently, studies of this cluster have come to conflicting conclusions regarding its dynamics and merger history. \citet{qrw96}, hereafter QRW, argue that A3266 is the result of a major merger of unequal mass clusters along the line of sight. The less massive cluster would have penetrated the main cluster from the southwest 4Gyr ago, with core collision around 1--2 Gyr ago. The secondary cluster passed through the main cluster forming a wedge-shaped tidal arm with velocities of 18,300~\kms \space north of the X-ray core \citep{mfg93}. Using 391 redshifts QRW estimate a velocity dispersion in the central region of 1400--1600~\kms, with a drastic fall to 700--800 \kms \space at 3~Mpc from the core. They noted that the velocity distribution is close to a Gaussian, although its spatial distribution is far from symmetric. \citet{fqw00} simulated the merger of both equal- and unequal-mass components with an N-body hydrodynamic model in which they include the evolution of dark matter. They concluded that a merger of similar-mass clusters would better resemble the observed galaxy distribution of their sample. \citet{rf00} presented an N-body hydrodynamic model in which the available spectroscopic data at that time was able to be reproduced for a merger an angle of about $45\degr$ to the line of sight and a mass-ratio of 5:2. However, they noted that there is a considerable flexibility in the allowed range of merger parameters and viewing angle.

\citet{hdd00} presented an alternative view. They posited that the morphology of A3266 was a consequence of a minor merger in the plane of the sky, where a small group of galaxies passed through the main cluster core from the southwest to the northeast. The X-ray temperature map shows evidence of shocked gas and the presence of radio galaxies southwest of the main cluster that can be explained as a result of ram-pressure stripping due to the merger \citep{ht02}. Using the pre- and post-shock gas temperatures, they calculated the relative gas velocity to be about 1400~\kms \space with a Gaussian velocity distribution. They agreed with a previous scenario of a merger of two clusters with a mass ratio of 4:1 \citep{rbl96}. \citet{sbp05} suggest that the observed X-ray shock wave region may be explained with two scenarios; the first in which a subcluster passed the main cluster core either from northeast to southwest 0.15--0.20 Gyr ago, or from southwest to northeast 0.8 Gyr ago where the subcluster is close to the turnaround point. In an XMM-Newton analysis of the cluster \citet{fhm06} detected a low-entropy region of gas near the cluster core and concluded that the existence of such a region must be due to the merger of a component with one 10th of the cluster mass in the plane of sky towards southwest.

Although most of the previous studies note the existence of substructure, there are some disagreements on the exact merger status of the cluster; explicitly, there is a lack of consensus about the phase of the core passage (pre- or post-merger) and the current direction of the cluster and subcluster motions. This paper is the first of a series in which we attempt to clarify the dynamical situation in A3266 using optical, radio and X-ray analyses. Here, we present 880 new spectroscopic redshifts in a $2\degr$-field around A3266. We combined these with existing values from the literature to build the most complete spectroscopic sample of A3266 to date with over 1300 measurements. With this much larger sample of redshifts, we determine the spatial and velocity distributions of the cluster and other structures in the region. This article is laid out as follows; Section~\ref{sec:data} describes the spectroscopic observations and sample construction. In Section~\ref{sec:analysis} the structure analysis of the cluster is explained. Section~\ref{sec:overall} depicts the overall cluster dynamics and possible merger scenarios. The summary is laid out in Section~\ref{sec:summary}. Throughout, we assume a standard Lambda-CDM model with $H_{0}=71$~km~s$^{-1}$~Mpc$^{-1}$, $\Omega_{m}=0.27$, and $\Omega_{\Lambda}=0.73$ which at the average redshift of A3266, $z=0.0594$, gives the scale of 1.132 kpc for $1\arcsec$.

\begin{table*}\caption{Details of the optical observations from which spectroscopic redshift data were obtained for this study.}
\begin{tabular}{ccccccc}
\hline
\multirow{2}{*}{Date} & Field & Number & b$_J$ Mag & T$_{exp}$ & $\lambda\lambda$ & $\Delta\lambda$ \\
 & Name & of Targets & Range & (s) & (\AA) & (\AA/pix) \\
\hline
97/12/29 & A3266--1 & 321 & 14.9--18.9 & 3$\times$1200 & 4785--7034 & 2.2 \\
97/12/29 & A3266--2 & 334 & 18.9--19.5 & 3$\times$1800 & 4785--7034 & 2.2 \\
99/10/05 & A3266--3 & 353 & 15.2--19.5 & 3$\times$900  & 3670--8080 & 4.3 \\
99/10/05 & A3266--4 & 334 & 19.5--20.0 & 3$\times$1500 & 3670--8080 & 4.3 \\
\hline
\end{tabular}\label{tab:opt_obs}
\end{table*}

\begin{table*}
\caption{Full data table for redshifts in a $2\degr$ region around A3266. Column 1 \& 2 give the coordinates of the objects, while columns 3-10 give the redshifts and their uncertainties from different datasets. The majority of the data are from this work (denoted by subscript `2dF'). Additional data were obtained from \citet{qrw96}, \citet{cbp09}, and various sources in the literature denoted by the subscripts `Q', `W', and `lit', respectively. The average redshift values and their corresponding uncertainties are given in column 11 \& 12. Column 13 provides a number corresponding to the source of the final redshift as follows: 1) this work, a value involving a new (2dF) measurement or an averaged value of two or more older datasets; 2) \citet{qrw96}; 3) \citet{cbp09}; 4) \citet{cz03}; 5) \citet{ggp90}; 6) \citet{qr90}; 7) \citet{gbt98}; 8) \citet{shn04}; 9) \citet{hmm12}. The first 10 rows are shown here and the full table of 1303 objects will be available at CDS via anonymous ftp to \url{cdsarc.u-strasbg.fr} (\url{130.79.128.5}) or via \url{http://cdsarc.u-strasbg.fr/viz-bin/qcat?J/MNRAS}.}
\label{tab:redshift}
\begin{tabular}{ccrrrrrrrrrrc}
\hline
RA & Dec & $cz_{\rm lit}$ & $\delta cz_{\rm lit}$ & $cz_{\rm W}$ & $\delta cz_{\rm W}$ & $cz_{\rm Q}$ & $\delta cz_{\rm Q}$ & $cz_{\rm 2dF}$ & $\delta cz_{\rm 2dF}$ & $\langle cz \rangle$ & $\delta \langle cz \rangle$ & \multirow{ 2}{*}{Ref} \\
\multicolumn{2}{c}{(J2000)} & \multicolumn{2}{c}{(\kms)} & \multicolumn{2}{c}{(\kms)} & \multicolumn{2}{c}{(\kms)} & \multicolumn{2}{c}{(\kms)} & \multicolumn{2}{c}{(\kms)} & \\
\hline
04:23:15.55 & -61:46:03.3 & --- & --- & ---  & --- & --- & --- & 17718  & 52  & 17718  & 52  & 1 \\
04:23:16.03 & -61:40:24.4 & --- & --- & ---  & --- & --- & --- & 22934  & 52  & 22934  & 52  & 1 \\
04:23:16.56 & -61:34:01.9 & --- & --- & ---  & --- & --- & --- & 57140  & 87  & 57140  & 87  & 1 \\
04:23:18.18 & -61:16:35.4 & --- & --- & ---  & --- & --- & --- & 37114  & 200 & 37114  & 200 & 1 \\
04:23:30.66 & -61:15:56.7 & --- & --- & ---  & --- & --- & --- & 19666  & 200 & 19666  & 200 & 1 \\
04:23:35.34 & -61:08:35.4 & --- & --- & ---  & --- & --- & --- & 17628  & 200 & 17628  & 200 & 1 \\
04:23:37.55 & -61:28:37.0 & --- & --- & ---  & --- & --- & --- & 83882  & 73  & 83882  & 73  & 1 \\
04:23:37.65 & -61:04:39.5 & --- & --- & ---  & --- & --- & --- & 17778  & 134 & 17778  & 134 & 1 \\
04:23:44.89 & -61:04:12.6 & --- & --- & ---  & --- & --- & --- & 17538  & 200 & 17538  & 200 & 1 \\
04:23:47.83 & -60:58:37.6 & --- & --- & ---  & --- & --- & --- & 48686  & 200 & 48686  & 200 & 1 \\
\hline
\end{tabular}
\end{table*}

\begin{figure}
\centering
{\includegraphics[width=2.4in]{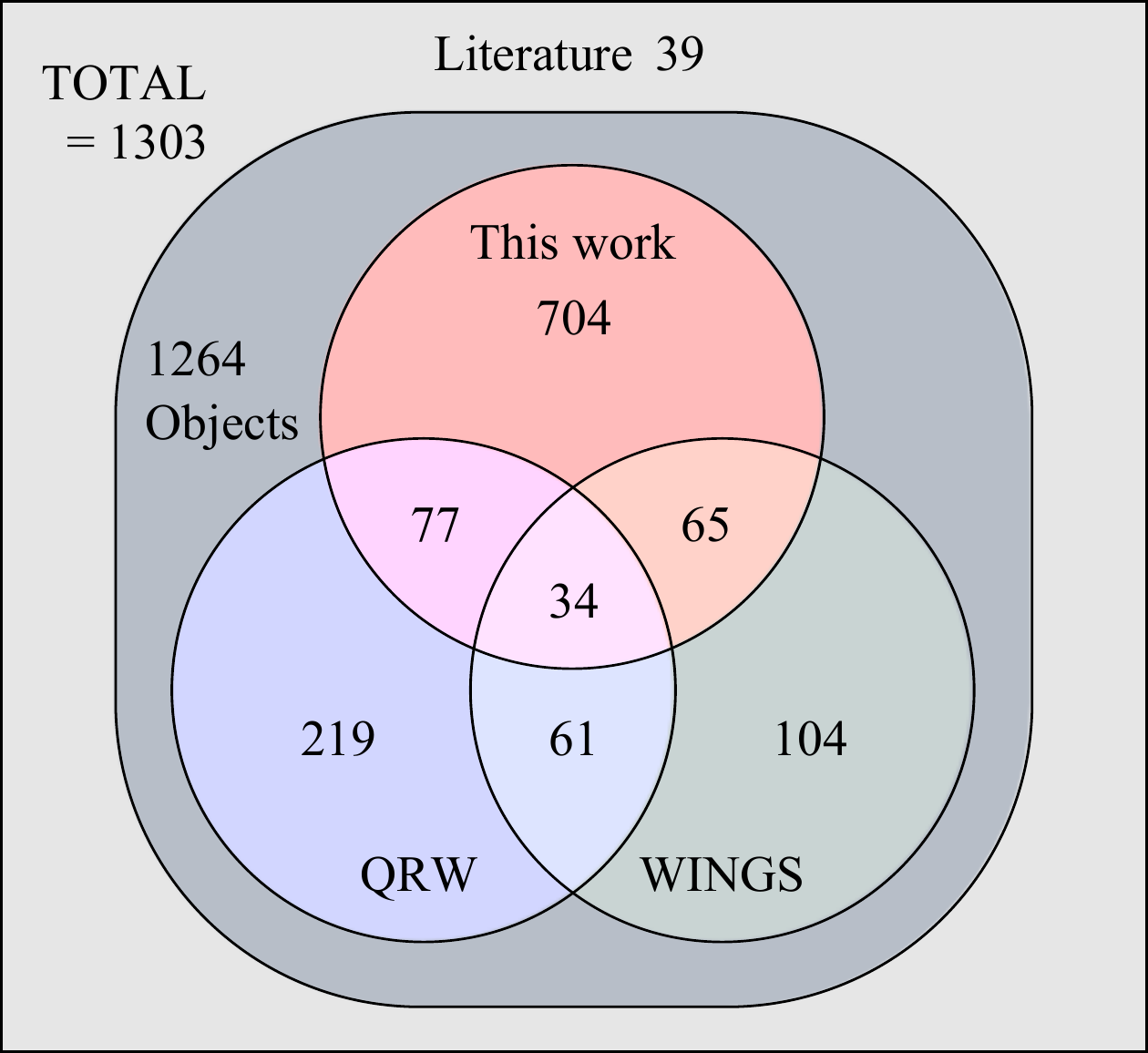}}
\caption{Venn diagram representing the compilation of spectroscopic redshifts in the vicinity of A3266 from 2dF, QRW, and WINGS datasets, and 39 unique objects extracted from the literature, summing up a total of 1303 objects.}\label{fig:diagram}
\end{figure}

\section{Spectroscopic Data}\label{sec:data}

Spectroscopic data were obtained for 880 galaxies observed with the 2dF instrument on the Anglo-Australian Telescope (AAT). The observations were performed on 29 December 1997 and 5 October 1999. Four fibre configurations were observed, all covering a single $2\degr$-diameter field around the nominal cluster centre of $\alpha$=04:31:09.7, $\delta$=$-$61:28:40 (J2000). Details of the observations are summarised in Table~\ref{tab:opt_obs}. The data were reduced with the {\tt 2dfdr} pipeline reduction software for 2dF\footnote{Software source code and full description available \url{https://www.aao.gov.au/science/software/2dfdr}}, and redshifts were measured using a modified version of the program used for the 2dF Galaxy Redshift Survey \citep{cdm01}.

\begin{figure*}
\centering
{\includegraphics[width=5.45in]{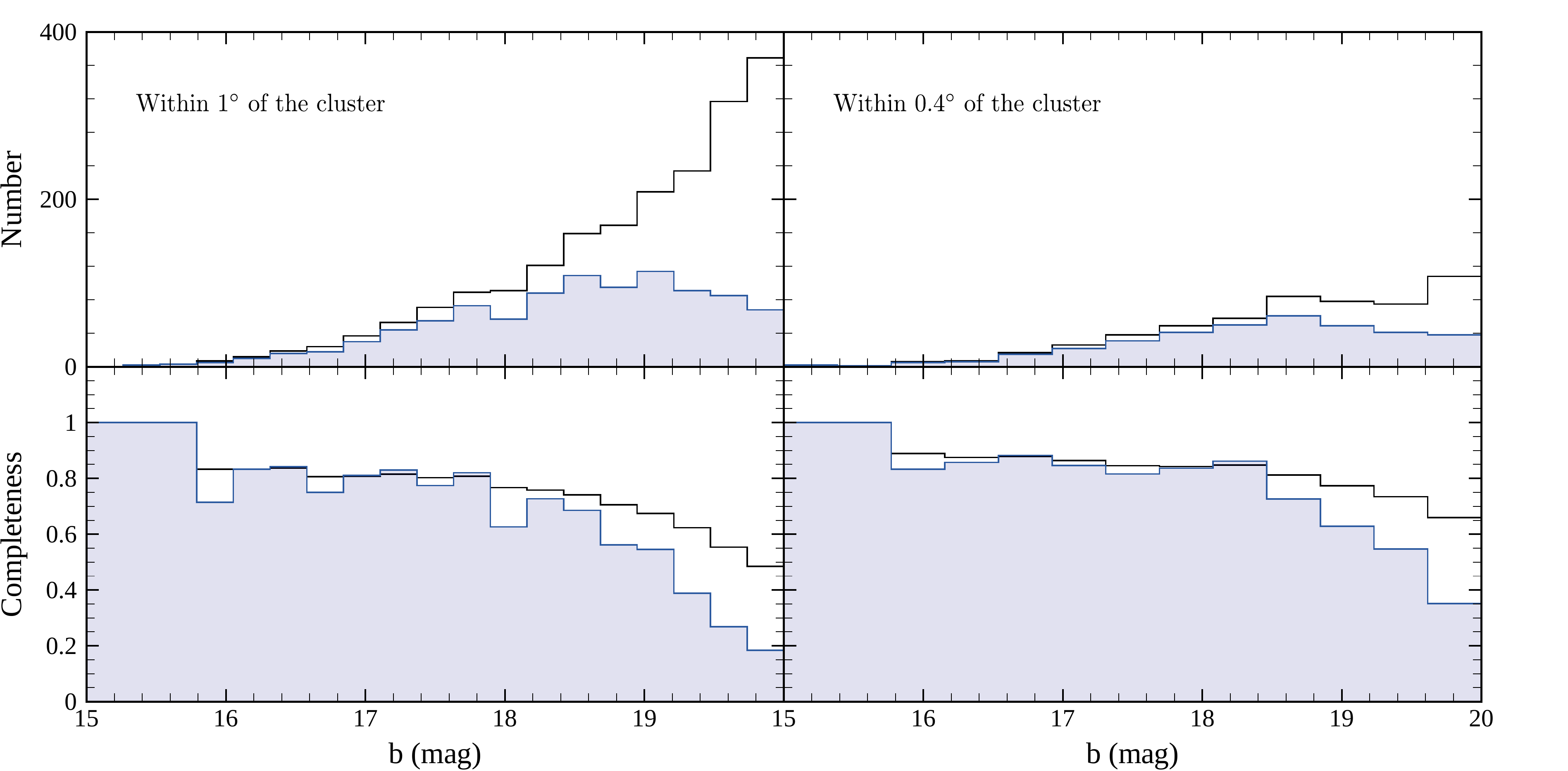}}
\caption{The completeness of the redshift sample within $1\degr$ (left panel) and $0.4\degr$ (right panel) of the cluster centre. The upper panel shows the differential number counts as a function of magnitude for all objects (open histogram) and objects with redshifts (blue filled). The lower panel shows the differential (blue filled) and cumulative (open) redshift completeness of the sample as a function of magnitude.}\label{fig:completeness}
\end{figure*}

\begin{figure*}
\centering
{\includegraphics[width=6.8in]{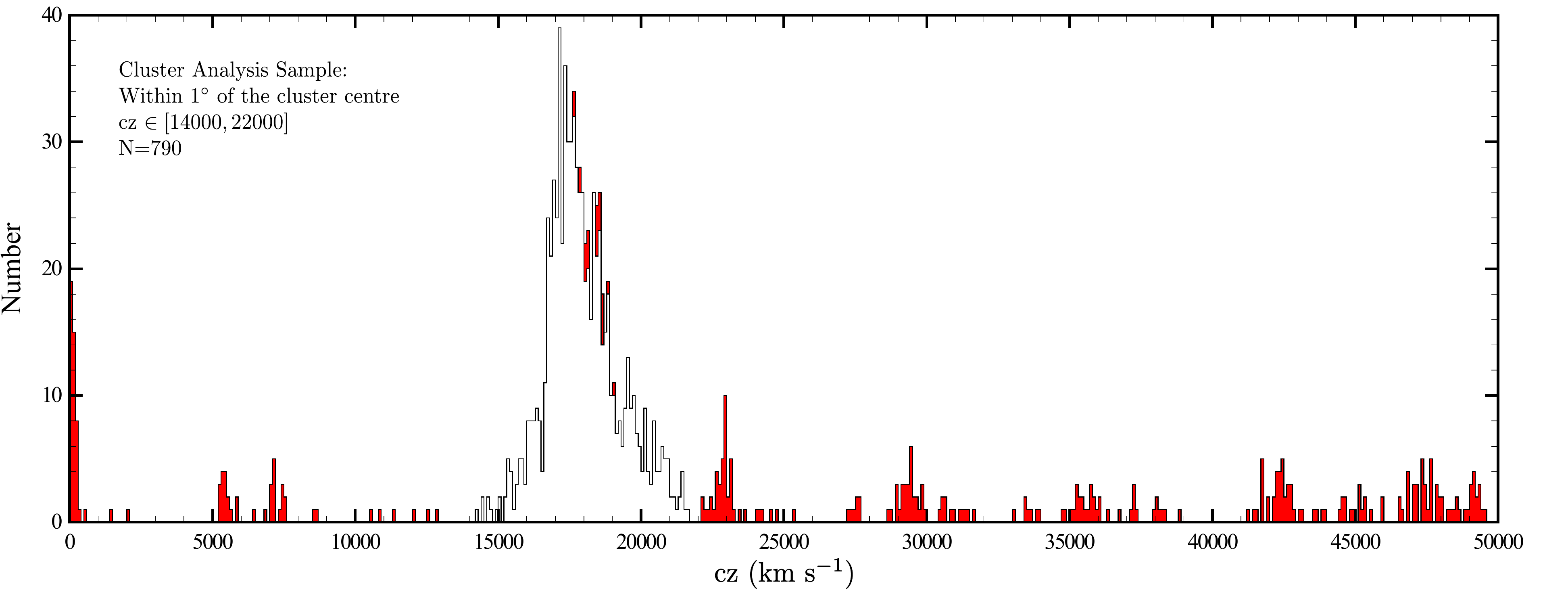}}
\caption{The distribution of galaxies in the full sample covering a $2\degr$-field is shown as a filled histogram. For clarity the velocity range of the plot has been truncated to 50,000~\kms. Overplotted is the cluster restricted sample (white) which covers a $1\degr$-field and a velocity range of 14,000--22,000~\kms.}\label{fig:histo}
\end{figure*}

\begin{figure}
\centering
{\includegraphics[width=\columnwidth]{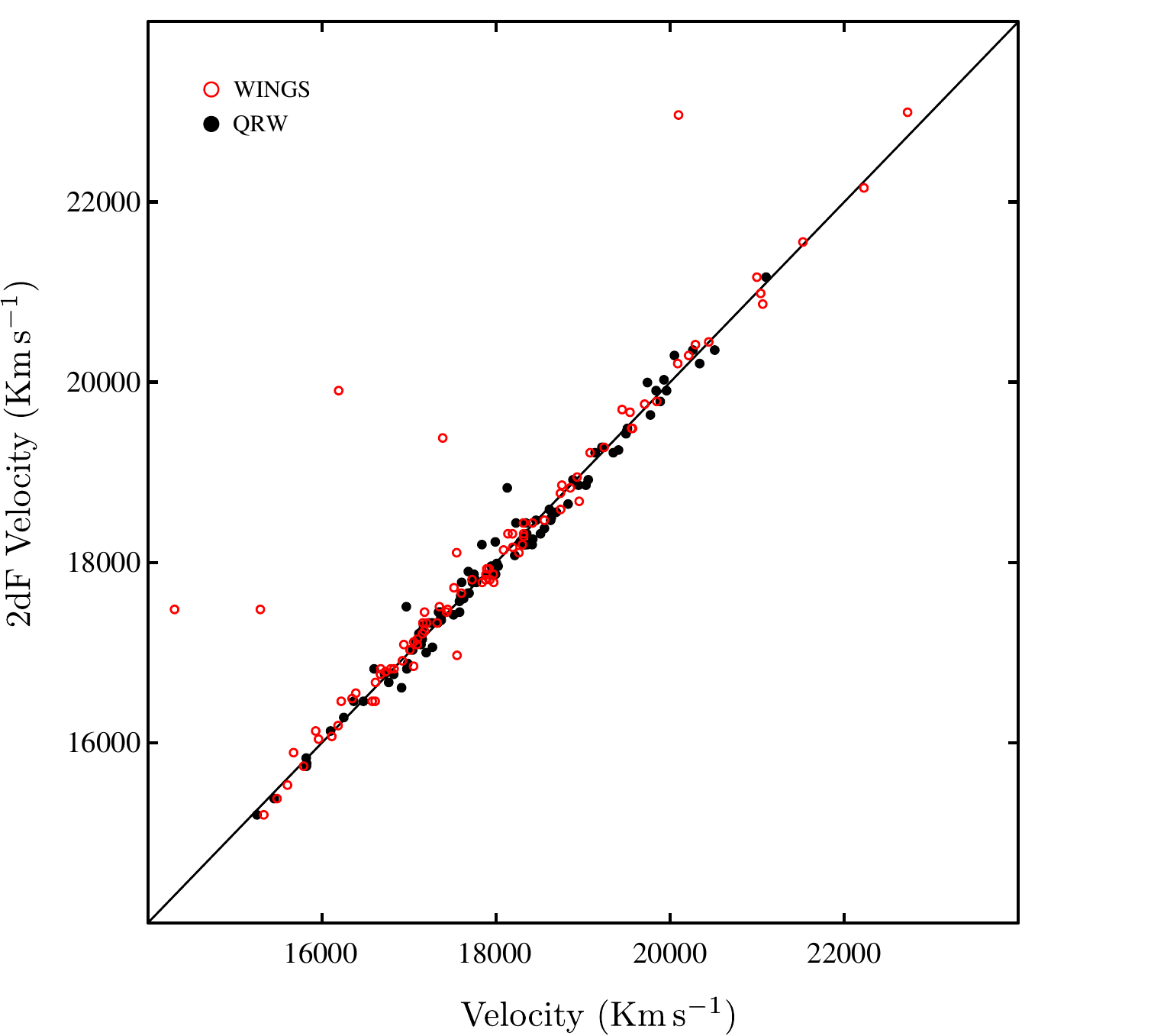}}
\caption{Comparison between the 2dF and WINGS (red open circles) and 2dF and QRW (black closed circles) velocities for objects measured in more than one survey. The lack line corresponds to the expected one-to-one fit between the surveys with no systematic differences.}\label{fig:1-1}
\end{figure}

Data from these observations were combined with available spectroscopic observations from the literature to produce the largest spectroscopic sample in the region of A3266 to date. Table~\ref{tab:redshift} lists the positions and redshifts for 1303 galaxies in the combined sample. The positions of the AAO catalogue come from the UK Schmidt Telescope southern sky survey plate for field F118, scanned with the Automated Plate Measuring machine \citep{mes90}. Redshifts are listed for 391 objects from the compilation of QRW, 880 objects from the previously unpublished 2dF data, 264 objects from the WIde-field Nearby Galaxy-cluster Survey \citetext{WINGS; \citealp{cbp09}}, and an additional 39 objects from various works in the literature within $1\degr$ of the nominal centre of A3266 (see Figure~\ref{fig:diagram}). Using a cross correlation radius of $4\arcsec$ there are 111, 99, and 95 objects in common between 2dF and QRW, 2dF and WINGS, and QRW and WINGS, respectively. Figure~\ref{fig:1-1} shows a comparison of 2dF redshifts to QRW and WINGS measurements. Using the iteratively re-weighted least squares fitting algorithm we find the relations $\mathrm{cz_{2dF}=0.99 \times cz_{QRW}}+150.95$\kms and $\mathrm{cz_{2dF}=1.00 \times cz_{QRW}}+41.94$\kms after 20 iterations. These relations show that 2dF redshifts are in good agreement with previous measurements and there are no systematic offsets between any of the catalogues used here. Excluding 39 objects for which the sets of redshift measurements differ by more than 3 times the combined error, the 2dF, QRW, and WINGS redshifts are consistent within their errors under a $\chi^2-$test. We have therefore used an error-weighted mean for the 198 objects with consistent redshift measurements. For the remaining 39 objects, 9 objects have common redshifts in the 2dF and QRW datasets, 11 are common to QRW and WINGS, 11 are common to WINGS and 2dF, a further 8 are common to all three datasets. We have re-inspected our data and the source of the QRW and WINGS redshifts as well as other redshift measurements in the literature to determine which values to adopt. In cases where there were a number of values in the literature which agreed with one of the 2dF, WINGS or QRW catalogues we adopted that catalogue value. Where there was no corroborating additional measurements we either performed an error-weighted average if the measurements were within 6 times the combined error, or alternatively, we selected the value with the lowest quoted error. For the 9 cases common to 2dF and QRW, in 6 cases we adopt the 2dF value, in 1 case the QRW value, and in 2 cases we use the error-weighted mean. For the 11 objects with inconsistent redshift measurements in the WINGS and QRW datasets we adopt the WINGS value for 5 cases, QRW value for 3 cases, and the error-weighted mean in 3 cases. For the 11 objects with inconsistent WINGS and 2dF values, we adopt WINGS, 2dF and error-weighted mean values for 4, 4, and 3 cases, respectively. For the remaining 8 objects with common redshift measurements in three datasets, in all cases, there are two consistent measurements; in 5 cases we use the error-weighted mean for WINGS and 2dF values, and in 3 cases QRW and 2dF values. The final combined dataset of 1303 objects is given in Table~\ref{tab:redshift}.

The completeness of the redshift samples restricted to objects within 1 and 0.4 degrees of the cluster centre are shown in Figure~\ref{fig:completeness} as a function of magnitude. The cumulative completeness drops to 80\% at $\rm{b_{j}}=18~\&~18.5$ within 1 and 0.4 degrees of the cluster centre, respectively, and is about 48\% and 66\% at $\rm{b_{j}}=20$ within the mentioned ranges. A histogram of the sample between 0 and 50,000~\kms \space is shown in Figure~\ref{fig:histo}. As can be seen in the figure, A3266 and possible merging groups lie in the velocity range of 14,000--22,000~\kms. In the rest of this paper the analysis sample is restricted to the 790 galaxies within $1\degr$ of the nominal cluster centre within the redshift range of 14,000--22,000~\kms\space (white part of the histogram), where the main population of galaxies corresponding to the cluster resides.

\begin{figure*}
\centering
{\includegraphics[width=\textwidth]{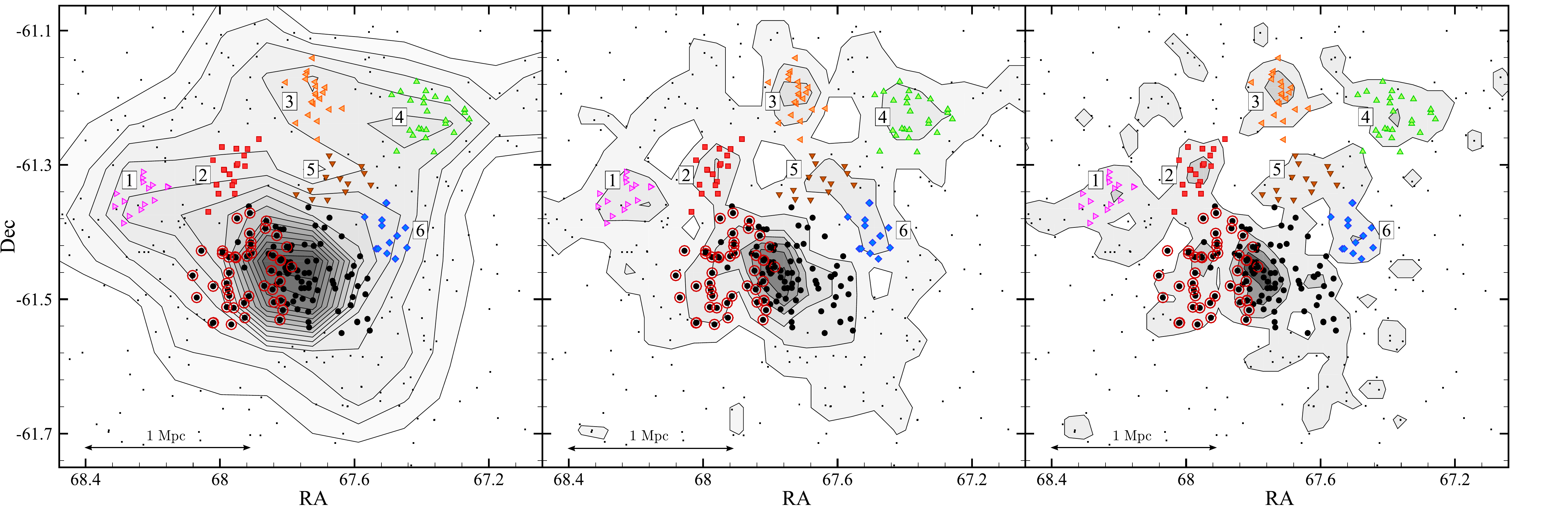}}
\caption[Spatial distribution of the structures surrounding the Abell 3266]{Spatial distribution of the structures identified by DBSCAN with an adopted $Eps$ value of 0.15~Mpc. Detected filaments and groups are numerically labelled, colour coded, and shown by different symbols. The cluster core is shown with filled black circles. The core has been decomposed into two subgroups via a 3D Lee-Fitchett test with the eastern core component denoted by red circles. Tiny black dots represent the non-assigned galaxies, i.e., noise points. All other colour coded symbols represent groups and filaments. The structures are overlaid on the galaxy number density maps of the cluster analysis sample, corresponding to cell sizes of 300, 150, and 100 kpc, from left to right, respectively. The scale line represents an angular extent 1~Mpc at the mean redshift of the cluster core.}\label{fig:density}
\end{figure*}

\begin{figure}
\centering
{\includegraphics[width=3in]{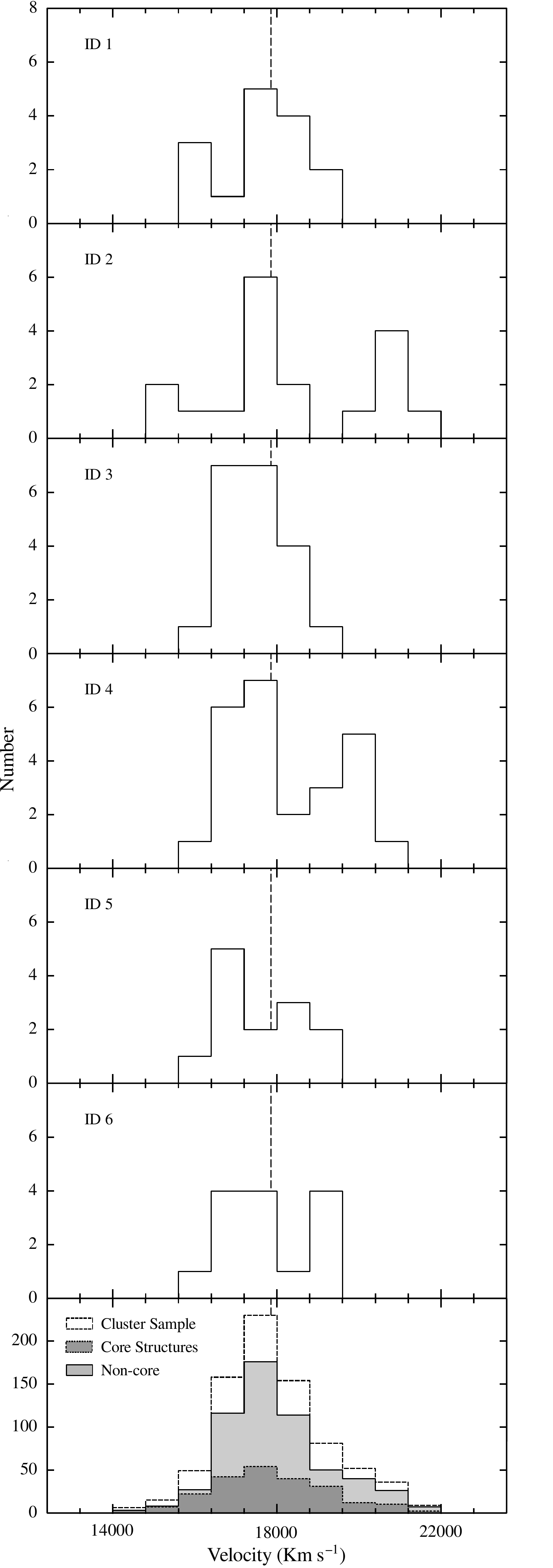}}
\caption[Velocity distributions of the groups and filaments surrounding the Abell 3266]{Velocity distributions of the structures shown in Figure~\ref{fig:density}. ID numbers correspond to the structure labels in Figure~\ref{fig:density}. Details of each structure are given in Table~\ref{tab:structures}. In all panels the vertical dashed line represents the (biweight) average velocity of the cluster sample.}\label{fig:histograms}
\end{figure}

\begin{table*}\caption{Details of the detected structures within A3266, including the cluster core and its substructures. The cluster core has been decomposed into two components via a Lee-Fitchett test denoted as western core component and eastern core component. The eastern core component is then further divided into two components, denoted here as the northwest substructure and the southeast substructure, see Section~\ref{sec:core} for further details of the decomposition of the core. Column 2 \& 3 give the average coordinates of the structure members, column 4 is the number of members of each structure, while column 5-10 provide the structure's average redshift and its error, and velocity dispersion and its uncertainty, relative to the CMB reference frame, respectively. Column 11 gives a class value which corresponds to 1) a filamentary structure or a structure consisting of a few groups aligned in the line of sight or 2) a galaxy group. Classification criteria for groups and filaments are based on those developed in \citet{dj14}.}\label{tab:structures}
\begin{tabular}{lccrccccrrc}
\hline
\multirow{2}{*}{Object} & RA                    & Dec & \multirow{2}{*}{N} & \multirow{2}{*}{$\overline{z}$} & \multirow{2}{*}{err} & $\overline{v}$       & err & $v_{d}$              & err & \multirow{2}{*}{Class} \\
                        & \multicolumn{2}{c}{(J2000)} &                    &                                 &                      & \multicolumn{2}{c}{(\kms)} & \multicolumn{2}{r}{(\kms)} &                        \\
\hline
Cluster Sample               & 04:31:10.38 & -61:24:27.3 & 790 & 0.0595 & 0.0002 & 17834 &  45 & 1209 &  39 & ... \\
ID: 1                        & 04:32:56.66 & -61:20:44.6 &  15 & 0.0592 & 0.0008 & 17751 & 253 &  877 & 146 & 1   \\
ID: 2                        & 04:31:51.96 & -61:18:27.3 &  18 & 0.0608 & 0.0018 & 18217 & 532 & 2043 & 276 & 1   \\
ID: 3                        & 04:30:52.71 & -61:11:51.7 &  20 & 0.0583 & 0.0005 & 17491 & 156 &  640 &  98 & 1   \\
ID: 4                        & 04:29:29.85 & -61:13:36.2 &  25 & 0.0602 & 0.0009 & 18051 & 280 & 1287 & 154 & 2   \\
ID: 5                        & 04:30:37.42 & -61:19:27.9 &  13 & 0.0585 & 0.0011 & 17534 & 331 & 1064 & 136 & 1   \\
ID: 6                        & 04:29:59.94 & -61:24:07.4 &  14 & 0.0592 & 0.0011 & 17744 & 330 & 1107 & 176 & 1   \\
Core                         & 04:31:12.25 & -61:27:45.6 & 118 & 0.0594 & 0.0005 & 17809 & 143 & 1462 &  99 & ... \\
Western Core Component (WCC) & 04:30:55.40 & -61:27:48.7 &  73 & 0.0609 & 0.0005 & 18261 & 164 & 1309 &  99 & ... \\
Eastern Core Component (ECC) & 04:31:38.62 & -61:27:40.8 &  45 & 0.0570 & 0.0008 & 17091 & 228 & 1449 & 115 & ... \\
Northwest Substructure (NWS) & 04:31:28.25 & -61:26:39.0 &  27 & 0.0552 & 0.0010 & 16553 & 288 & 1361 & 145 & ... \\
Southeast Substructure (SES) & 04:31:58.34 & -61:29:08.2 &  18 & 0.0600 & 0.0010 & 17984 & 307 & 1179 & 216 & ... \\
\hline
\end{tabular}
\end{table*}

\section{Structure analysis of the Abell 3266}\label{sec:analysis}

In order to examine substructure in the field the Density-Based Spatial Clustering of Applications with Noise (DBSCAN) algorithm was used \citep{eks96}. Typically DBSCAN has not been used in astronomy, but it has recently been shown to be particularly good at detecting structures in spectroscopic samples such as the process used recently for structure analysis in the Chandra Deep Field-South \citep{dj14} and for the massive cluster A3888 \citep{sjd16}. DBSCAN is a friend-of-friend clustering method that uses two input parameters, the neighbouring distance $(Eps)$ and a minimum number of points $(MinPts)$ which determines the detection threshold for objects that should be considered to be grouped. The algorithm counts the number of objects $(N)$ within the $Eps$-radius of the object of interest $(p)$. If the condition, $N > MinPts$, is satisfied, $p$ will be labelled as a `core point'. Objects that are not core points but are within the neighbourhood of a core point are labelled as `border points'. All the other points are discarded from the rest of the analysis. Afterwards, the algorithm recursively evaluates which of the core or border points are `connected', or within the $Eps$-radius of each other. All the connected core and border points are grouped as a single structure, and the algorithm moves on to the next unassigned point until there are no unevaluated points left. The number of input parameters can be reduced to one, only $MinPts$, using the so-called sorted $k$-dist graph. The sorted $k$-dist plot is made by calculating and sorting the distances of sample objects to their k-nearest object in a descending order, where $k=MinPts-1$. The $Eps$ value is determined where a significant turnover or sudden change in the slope of the graph is observed. It has been shown that the results of the algorithm are not particularly sensitive to the value of $MinPts$ (see \citealp{eks96} for more details regarding the parameter selection process).

Although DBSCAN can be applied to three-dimensional data by defining a pseudo-distance parameter, we only use two-dimensional projected data in order to avoid making assumptions on the velocity distribution of structures. This is due to the weakness of DBSCAN in concurrent detection of structures with substantially different densities in one or more dimensions, e.g., radial filaments and compact groups. The detected structure can be further investigated in velocity space to determine their properties and classification. Here the input parameter adjustment was performed by choosing $MinPts=6$, a sorted 5-dist plot of the sample was generated and $Eps=0.15$~Mpc was found and adopted for further analysis. Subsequently, the DBSCAN algorithm was applied to the spatial and velocity restricted catalogue to determine the structures surrounding Abell 3266. Based on spatial distribution, the overall population was divided into six groups and filaments, along with the cluster core.

\begin{figure*}
\centering
{\hspace{-1.2cm}\includegraphics[width=5.6in]{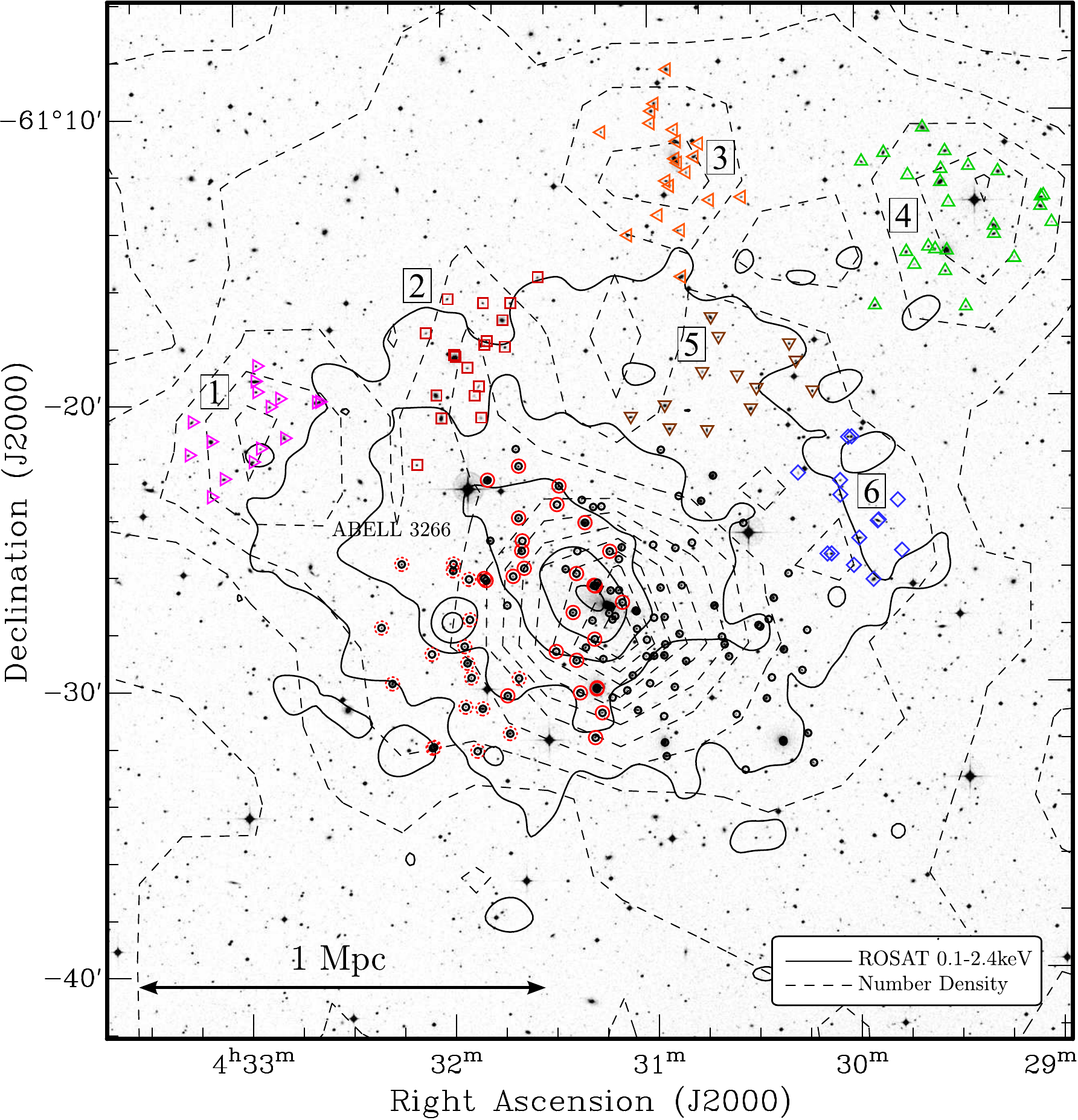}}
\caption[Multi-wavelength image of the Abell 3266]{The spatial distributions of the detected structures determined by DBSCAN are overlaid on a DSS optical image of the Abell 3266. The legends are the same as Figure~\ref{fig:density}, except here the eastern core component has been further decomposed into two groups denoted by solid and dashed lines (see Section~\ref{sec:core}). The ROSAT PSPC soft-band X-ray (0.1-2.4 keV) intensity map of the cluster, convolved with a $50\arcsec$ Gaussian kernel, is shown with black contours. The galaxy number density map of the cluster analysis sample is shown by black dashed contours. The density contours start at the local $3\sigma_{\textrm{rms}}$ level and then increase by steps of 11 galaxies per Mpc$^{2}$. The scale line represents an angular extent of 1~Mpc at the mean redshift of the cluster core.}\label{fig:allin}
\end{figure*}

The spatial distribution of the detected structures are shown in Figure~\ref{fig:density}. The identified groups and filaments are labelled and shown with coloured geometrical symbols. In addition, the cluster core is represented by black circles. A further analysis on the core was undertaken to determine substructure in the cluster population by performing a three-dimensional\footnote{3-dimensional refers to analysis of spatial and line-of-sight velocity data points, concurrently.} Lee-Fitchett test \citetext{hereafter Lee3D, \citealp{l79}; \citealp{fw87}; \citealp{f88}}. The core was decomposed into two components, split in both the spatial and velocity domains, with one component located in the east at a slightly lower than average redshift and one component in the west with slightly higher than average redshift (see Section~\ref{sec:core}). The eastern core component is denoted here by red circles around the black points marking core galaxies. The non-assigned galaxies are shown with tiny black dots across the field. These structures are overlaid on surface density maps of the cluster region. The panels in Figure~\ref{fig:density} from left to right are the density contours based on bilinear interpolation of galaxy number counts within the cell sizes with constant extent of 300, 150, and 100 kpc, respectively. Each set of density contours start at about local $3\sigma_{\textrm{rms}}$ level and then increase by steps of 11, 44, 100 galaxies per Mpc$^{2}$, and represent high sensitivity \& low resolution to low sensitivity \& high resolution density map of the cluster, respectively. The scale line in the figure represents an angular extent of 1~Mpc at the mean redshift of the cluster core $(z=0.0594)$. In addition, we represent the velocity distributions of the detected groups and filaments, and the cluster core by histograms with bin width of 800~\kms in Figure~\ref{fig:histograms}. ID numbers in Figure~\ref{fig:histograms} correspond to labelled structures in Figure~\ref{fig:density}. We report the properties of the detected structures, inferring the redshift location and rest-frame velocity dispersion along with their 68 per cent confidence intervals\footnote{Calculated by applying the biweight, gapper, and jackknife methods \citep*{bfg90}.}, relative to the reference frame of the Cosmic Microwave Background (CMB, \citealp{fcg96}) in Table~\ref{tab:structures}.

\subsection{Structures Surrounding the Abell 3266 Core}

\begin{figure}
\centering
\includemedia
[
label=Cluster3D,
width=\linewidth,
height=\linewidth,
add3Djscript=Figures/3D_Plot/Initialization.js,
deactivate=pageinvisible,
3Dviews=Figures/3D_Plot/Cluster3D.vws,
3Dmenu
]
{\includegraphics{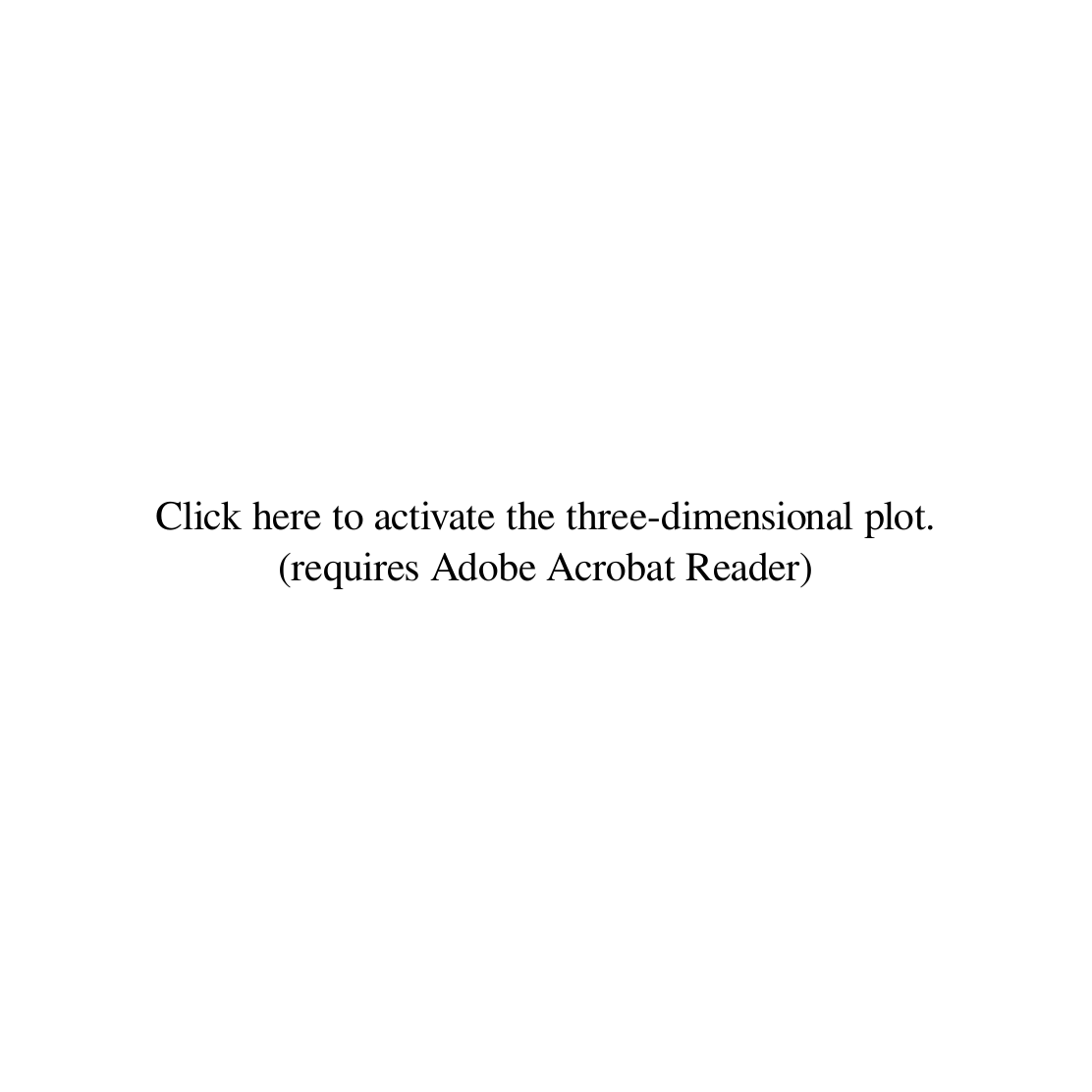}}{Figures/3D_Plot/Cluster.prc}
\mediabutton[
  overface={\includegraphics[height=1.5em]{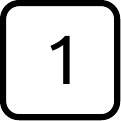}},
  downface={\includegraphics[height=1.5em]{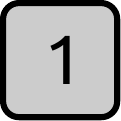}},
  jsaction=Cluster3D:{
    for(var i=792; i<807; i++) {
      annotRM['Cluster3D'].context3D.toggleNodeByID(i);
    }
    annotRM['Cluster3D'].context3D.toggleNodeByID(1119);
  }
]{\includegraphics[height=1.5em]{Figures/3D_Plot/Buttons/1_1.pdf}}
\mediabutton[
  overface={\includegraphics[height=1.5em]{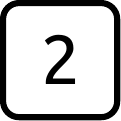}},
  downface={\includegraphics[height=1.5em]{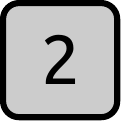}},
  jsaction=Cluster3D:{
    for(var i=807; i<825; i++) {
      annotRM['Cluster3D'].context3D.toggleNodeByID(i);
    }
    annotRM['Cluster3D'].context3D.toggleNodeByID(1120);
  }  
]{\includegraphics[height=1.5em]{Figures/3D_Plot/Buttons/2_1.pdf}}
\mediabutton[
  overface={\includegraphics[height=1.5em]{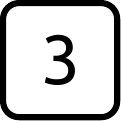}},
  downface={\includegraphics[height=1.5em]{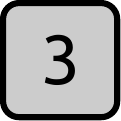}},
  jsaction=Cluster3D:{
    for(var i=825; i<845; i++) {
      annotRM['Cluster3D'].context3D.toggleNodeByID(i);
    }
    annotRM['Cluster3D'].context3D.toggleNodeByID(1121);
  }  
]{\includegraphics[height=1.5em]{Figures/3D_Plot/Buttons/3_1.pdf}}
\mediabutton[
  overface={\includegraphics[height=1.5em]{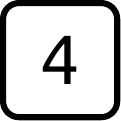}},
  downface={\includegraphics[height=1.5em]{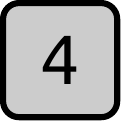}},
  jsaction=Cluster3D:{
    for(var i=845; i<870; i++) {
      annotRM['Cluster3D'].context3D.toggleNodeByID(i);
    }
    annotRM['Cluster3D'].context3D.toggleNodeByID(1122);
  }  
]{\includegraphics[height=1.5em]{Figures/3D_Plot/Buttons/4_1.pdf}}
\mediabutton[
  overface={\includegraphics[height=1.5em]{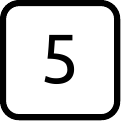}},
  downface={\includegraphics[height=1.5em]{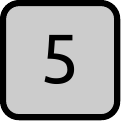}},
  jsaction=Cluster3D:{
    for(var i=870; i<883; i++) {
      annotRM['Cluster3D'].context3D.toggleNodeByID(i);
    }
    annotRM['Cluster3D'].context3D.toggleNodeByID(1123);
  }  
]{\includegraphics[height=1.5em]{Figures/3D_Plot/Buttons/5_1.pdf}}
\mediabutton[
  overface={\includegraphics[height=1.5em]{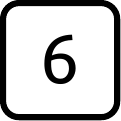}},
  downface={\includegraphics[height=1.5em]{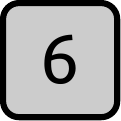}},
  jsaction=Cluster3D:{
    for(var i=883; i<897; i++) {
      annotRM['Cluster3D'].context3D.toggleNodeByID(i);
    }
    annotRM['Cluster3D'].context3D.toggleNodeByID(1124);
  }  
]{\includegraphics[height=1.5em]{Figures/3D_Plot/Buttons/6_1.pdf}}
\mediabutton[
  overface={\includegraphics[height=1.5em]{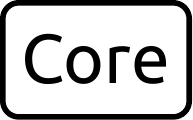}},
  downface={\includegraphics[height=1.5em]{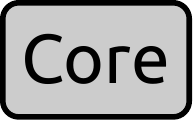}},
  jsaction=Cluster3D:{
    for(var i=897; i<1015; i++) {
      annotRM['Cluster3D'].context3D.toggleNodeByID(i);
    }
    annotRM['Cluster3D'].context3D.toggleNodeByID(1118);
  }  
]{\includegraphics[height=1.5em]{Figures/3D_Plot/Buttons/Core_1.pdf}}
\mediabutton[
  overface={\includegraphics[height=1.5em]{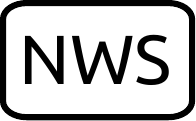}},
  downface={\includegraphics[height=1.5em]{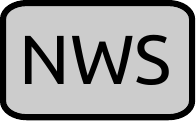}},
  jsaction=Cluster3D:{
    for(var i=1028; i<1082; i++) {
      annotRM['Cluster3D'].context3D.toggleNodeByID(i);
    }
  }  
]{\includegraphics[height=1.5em]{Figures/3D_Plot/Buttons/NWS_1.pdf}}
\mediabutton[
  overface={\includegraphics[height=1.5em]{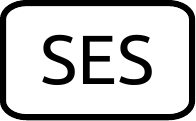}},
  downface={\includegraphics[height=1.5em]{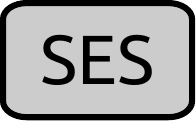}},
  jsaction=Cluster3D:{
    for(var i=1082; i<1118; i++) {
      annotRM['Cluster3D'].context3D.toggleNodeByID(i);
    }
  }  
]{\includegraphics[height=1.5em]{Figures/3D_Plot/Buttons/SES_1.pdf}}
\mediabutton[
  overface={\includegraphics[height=1.5em]{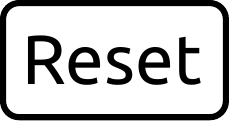}},
  downface={\includegraphics[height=1.5em]{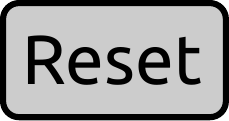}},
  3Dgotoview=Cluster3D:Default
]{\includegraphics[height=1.5em]{Figures/3D_Plot/Buttons/Reset_1.pdf}}
\caption[Three-dimensional visualisation of the A3266 cluster sample.]{Three-dimensional visualisation of the A3266 cluster sample. ID numbers correspond to the structure labels in Figure~\ref{fig:density}. The cluster core and its eastern core component split into two parts being the Southeast Subtructure (SES) and Northwest Subtructure (NWS) are shown with black spheres, and magenta and purple circles, respectively. Structures may be hidden or shown using their corresponding toggles located at the bottom of the plot. (This 3D plot can be displayed and interacted using Adobe Acrobat Reader v8.0 or higher. The position and viewpoint of the camera can be changed using the mouse, e.g., drag with the left mouse button. Additionally a menu for interactive commands is available via the right mouse button.)}\label{fig:3d}
\end{figure}

In Figure~\ref{fig:allin} we represent the detected structures along with the cluster core overlaid on the DSS optical image of the Abell 3266 cluster. The legend of the structures is the same as Figure~\ref{fig:density}. In addition, the ROSAT PSPC soft band [0.1-2.4 keV] X-ray intensity map \citep{vab99} is represented by black contour lines (convolved with a FWHM of $50\arcsec$). Furthermore the galaxy number density map of the cluster analysis sample (corresponding to cell sizes of 300 kpc), is shown by black dashed lines. An interactive three-dimensional (RA, Dec, and velocity) visualisation of the cluster sample and its associated structures is represented in Figure~\ref{fig:3d}. This plot was conducted with the S2PLOT programming library \citep{bfb06} and can be activated and interacted with using Acrobat Reader v8.0 or higher. We now concentrate on properties of the detected structures surrounding the cluster core:

\textbf{Structures 1 \& 2} classified as filaments or a series of groups along the line of sight in the classification scheme of \citet{dj14}. Both structures have a very broad velocity dispersion ($v_{d}=877^{+146}_{-146}$ \& $2043^{+276}_{-276}$~\kms, respectively), typical of radial filamentary structures. A further inspection of Figure~\ref{fig:3d} shows that Structures 1 has a filamentary morphology both in its redshift and spatial distributions. However, Structure 2 contains two major concentration of galaxies, each of which with five galaxies, located in the line of sight at $z=0.0591\pm0.0002$ \& $0.0700\pm0.0002$ with $v_{d}=132^{+83}_{-27}$ \& $173^{+115}_{-37}$, respectively. Structure 2 appears to be associated with the low entropy region found by \citet{fhm06}.

\textbf{Structure 3} is an isolated group located at distance of 1~Mpc to the north of the cluster centre. Structure 3 with 20 galaxy members has a symmetrical redshift distribution with a mean redshift of $z=0.0583\pm0.0005$ and a velocity dispersion of $v_{d}=640^{+98}_{-98}$. Despite the velocity dispersion of Structure 3, which is sufficiently high for classification as a cluster \citep{sr91}, there is no extended X-ray emission visible in the ROSAT 0.1-2.4 Kev image \citep{vab99} of the area (see Figure~\ref{fig:allin}). In order to determine whether Structure 3 is a group with its own substructure, and therefore has a high velocity dispersion, or an under-luminous X-ray cluster, we estimated the sampling rate of the structure if our redshift sample was complete up to absolute magnitude of $M_c=-15$. The estimation was made based on a Schechter luminosity function,
\begin{equation}
\phi(M)=k\phi^{*}10^{0.4(M^*-M)(\alpha+1)}\exp(-10^{0.4(M^*-M)}),
\end{equation}
where we take the parameters $M^*=-22.5$ and $\alpha=-1.25$ \citep{rab04}. The absolute magnitude of an object as a function of redshift, $M_l$, is calculated by using the apparent magnitude limit of our redshift sample, I = 17. We also apply a K-correction to this absolute magnitude, assuming that the object is an elliptical galaxy at $z=0.06$. The normalization parameter, $\phi^{*}$, was calculated by solving the equation,
\begin{equation}
\int_{-\infty}^{M_l} \phi(M) dM = N_r,
\end{equation}
where $N_r=20$ is the structure's sampling rate. Using the mentioned parameters we estimate that Structure 3 must have at most $\sim$ 140 galaxies with absolute magnitudes smaller than magnitude cut-off of $M_c=-15$. We therefore classify this structure as a large group or a poor cluster associated with A3266 with the lack of X-ray emission making the group scenario more likely.

\textbf{Structure 4} has two major peaks in its velocity distribution (see Figure~\ref{fig:histograms}). These peaks in the velocity distribution correspond to two populations of galaxies, each of which with two components, detected as a whole by DBSCAN (see Figure~\ref{fig:3d}). The northern substructure is located in background of the cluster at $z=0.0655\pm0.0007$, whereas the southern one is in foreground of the cluster at $z=0.0578\pm0.0005$.\footnote{However, this may not be the case, if we consider the dynamics of a merger process. The problem arises, since it is usually difficult to accurately separate the Hubble flow from the peculiar velocity; the northern substructure could be closer to the observer, yet, it may have a positive peculiar velocity due to the gravitational force exerted by the southern substructure. This results in an inward motion direction and a higher redshift in the observer's viewpoint. Likewise, the southern structure may be located in foreground of the northern substructure.} Both northern and southern substructures have relatively high velocity dispersion ($v_{d}=561^{+104}_{-104}$ and $v_{d}=477^{+111}_{-111}$~\kms, respectively), as expected for gravitationally disturbed structures.

\textbf{Structure 5 \& 6} build up a semicircular filament located at the northwest of the cluster (see last panel of Figure~\ref{fig:density}). Structure 5 consists of two populations with 8 and 5 galaxies along the line of sight, located at $z=0.0565\pm0.0005$ \& $0.0625\pm0.0004$, with velocity dispersions of $v_{d}=404^{+117}_{-117}$ \& $227^{+151}_{-49}$~\kms, respectively. Structure 6 which is located in the west of the cluster core, has filamentary morphology both in the velocity and spatial distributions.

\subsection{The Core of Abell 3266}\label{sec:core}

In the DBSCAN analysis, a population of 118 galaxies in the core region were separated from surrounding groups and filaments. This population represents an elongated structure aligned with the X-ray gas morphology, which is typically found in the clusters in a merger process. To detect possible substructure in this population of galaxies the Lee3D test was used. The Lee3D test is based on a maximum likelihood technique, in which a statistical significance is assigned to various clustering hypotheses. A hypothesis is made based on a rotating line passing through the cluster and separating the galaxies into two groups. For any given hypothesis, which corresponds to a division line with different 3D orientation, the likelihood ratio criterion is calculated and the hypothesis maximizing the ratio is chosen for optimized separation. The test is one of the most effective multivariate clustering methods developed specifically for structures with only one significant substructure \citep{prb96}.

\begin{figure}
\centering
{\includegraphics[width=\columnwidth]{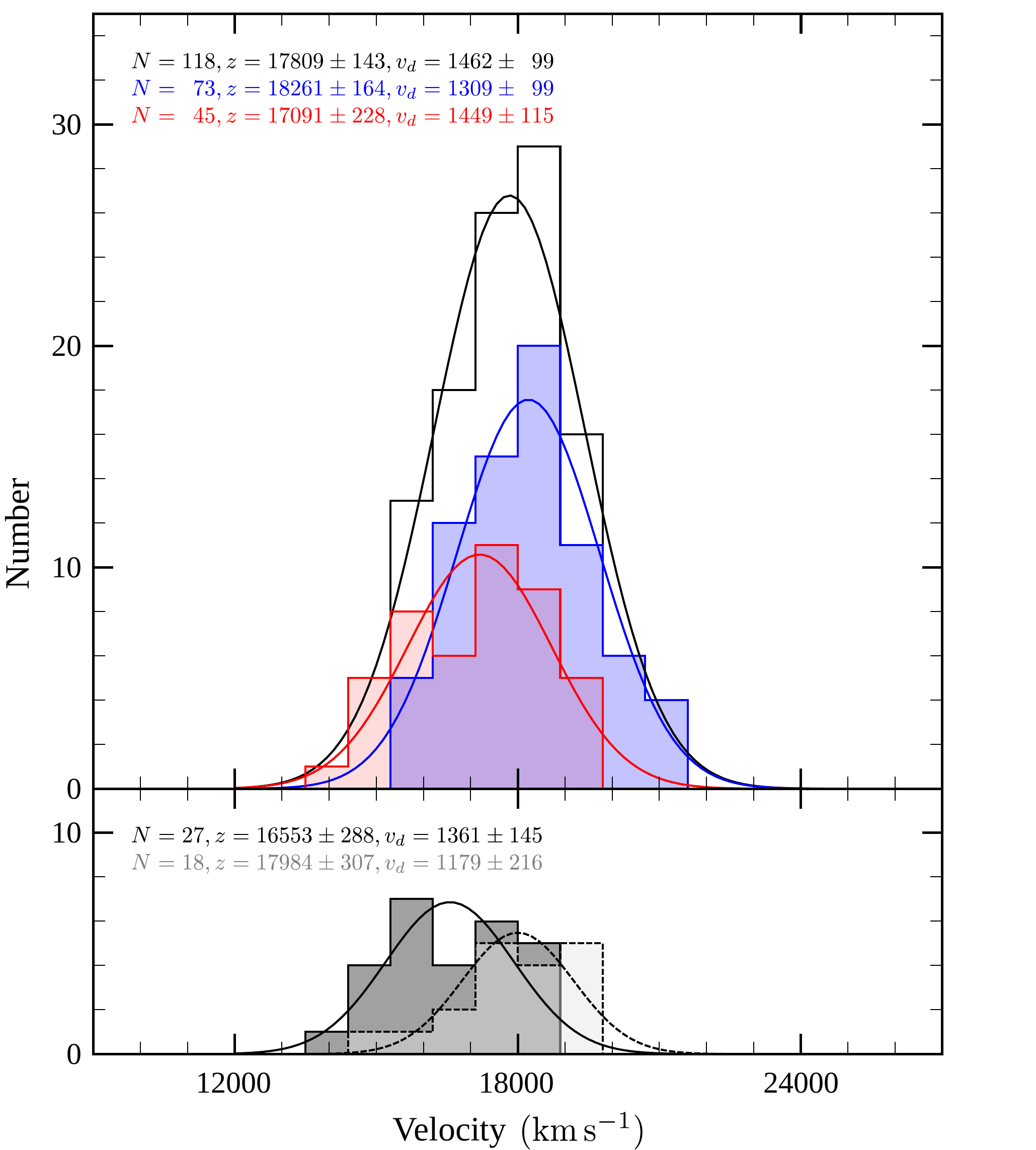}}
\caption{Velocity distribution of the cluster core and its decomposition into two substructures (top panel) and components of the ECC (bottom panel). Black, red, and blue histograms and Gaussian curves represent the velocity distributions of the whole core population (shown with black circles in Figures~\ref{fig:density}~\&~\ref{fig:allin}), the ECC (shown with red circles in Figures~\ref{fig:density}~\&~\ref{fig:allin}), and the WCC, respectively. The lighter shade histogram in the bottom panel corresponds to the further decomposition of the ECC into two component parts, NWS and SES, shown with red solid and dashed circles in Figure~\ref{fig:allin} and purple and magenta circles in Figure~\ref{fig:3d}.}\label{fig:histocore}
\end{figure}

The Lee3D test was applied to a sample of 118 galaxy members using the {\tt CALYPSO} package (Dehghan \& Johnston-Hollitt in prep), and it successfully decomposed the cluster into two galaxy distributions. One is a moderately significant subgroup with 45 galaxies to the east of the cluster core with slightly lower redshift than the cluster core average. The remaining 73 galaxies are the dominant cluster core to the west with a slightly higher than average redshift. We refer to these components as the East Core Component (ECC) and the West Core Component (WCC) in Table~\ref{tab:structures}. The significance of the subgrouping was calibrated by performing a Monte Carlo (MC) simulation of 800 distributions under the null hypothesis of the form $\sigma=\sigma_{0}(r^2+r_c^2)^{-1/2}$, where $\sigma_{0}$ is a constant value and $r_c=0.001$ corresponds to the assumption that galaxy clusters have power law surface density profile (see \citealp{bt86,f88} for more details). The null hypothesis rejection rate was $98.75\% \pm 1.15\%$ ($99\%$ confidence intervals). The ECC is shown with red circles in Figures~\ref{fig:density}~\&~\ref{fig:allin}. The top panel of Figure~\ref{fig:histocore} shows the velocity distribution of the whole cluster core, ECC, and WCC with black, red, and blue histograms and their corresponding Gaussian fits.

The mean redshift of the ECC, relative to the 3K background, is $z=0.0570\pm0.0008$, which is slightly different from that of the whole cluster core ($z=0.0594\pm0.0005$). The morphology of the ECC is considerably scattered, whereas the WCC remains relatively compact. Additionally, the velocity distribution of the ECC includes two peaks (red histogram in Figure~\ref{fig:histocore}). We performed a secondary Lee3D test on the 45 galaxies of the ECC to investigate whether it can be further decomposed. Subsequently, this subgroup was further separated into two substructures with 18 \& 27 members located at $z=0.0552\pm0.0010$ \& $z=0.0600\pm0.0010$. The statistical significance of the separation was $99.79\% \pm 0.13\%$ which was calculated by MC simulation of 8000 samples, this time with $r_c=0.5$ (core model in \citealp{f88}) due to the smaller extent of the ECC. The smaller substructure of the ECC resides to the southeast on the edge of the diffuse X-ray emission, whereas, the larger component is almost entirely associated with both the X-ray elongation and the low entropy X-ray gas detected by \citet{fhm06}. We refer to these structures as Southeast Subtructure (SES) and Northwest Subtructure (NWS). Thus following a second stage Lee3D analysis the core of 118 galaxies can either be considered to be bimodal consisting of the ECC and WCC, or trimodal consisting of the WCC, NWS, and SES. Taken together these results suggest that what is now the northwestern component of the ECC has merged with the cluster either from the southwest or northeast. Examining the Lee3D results suggest the merger axis is about $120\degr$ clockwise from North. The WCC with 73 members, which would be the residual cluster structure following the merger, was found to have $z=0.0609\pm0.0005$, and a velocity dispersion of $v_{d}=1309^{+99}_{-99}$.

\section{Overall Cluster Dynamical Picture}\label{sec:overall}

From the Lee-Fitchett analyses we find that the core is decomposed into two major components with a mass-ratio of roughly 1:2, of which the ECC further breaks up into two parts. Since the majority of the galaxies in the SES are located on the edge of the X-ray emission, this structure appears to be unrelated to the main merger event. The position of the remaining two subgroups, with a mass-ratio of 1:3, supports two possible scenarios of an ongoing merger in the plane of sky along a southwest-northeast axis. The first scenario is suggestive of a past core passage merger with the NWS having come from the southwest, moved to the northeast, and stripped the cluster ICM during the passage. This interpretation can explain the relatively large dispersion of NWS and extension of the hot diffuse gas to the northeast of the core, as well as the offset between peak positions of the X-ray emission and surface density or the location of the Brightest Cluster Galaxy (BCG). \citet{sbp05} considered a similar scenario and found that observed properties of the X-ray gas can only be explained if the mass-ratio for the merger is considerably larger than 1:3.

According to the second scenario the NWS has entered the cluster from northeast and is on its way to pass the core and exit from the southwest, during which the low-entropy gas of this smaller group has been stripped due to the high density ICM of the WCC. This would appear to be consistent with the detection of the low entropy tail, with the precise shape of that seen in the XMM analysis, and satisfies the prediction of an infalling subcluster group, as outlined in \citet{fhm06}. Considering a similar situation for a merger from the southwest, \citet{sbp05} performed a numerical simulation of a merger with a mass-ratio of 1:3. About 0.2~Gyr after the first core passage, they found a morphology that could explain most of the X-ray properties of A3266. This mass-ratio is close to what we measure, however, our optical analysis suggests that the core passage has not happened at this stage as the western edge of the NWS is almost aligned with the position of the BCG. In addition, the NWS appears to be excessively disturbed for a structure in a pre-merger state. Overall, it appears that optical and X-ray analyses, alone, are not sufficient for definitive decision on the dynamics of A3266, which seems more complex than a simple merger process.

Furthermore, the cluster's total gas mass ($M_{500}=1.2 \times 10^{14} M_{\odot}$, \citealp{fhm06}) corresponds to a velocity dispersion of about 540~\kms, according to the empirical mass-velocity relation \citep{v05}, which is considerably lower than velocity dispersion we measure ($v_d \sim 1400$\kms), implying a significantly disrupted system. In addition, the cluster core is surrounded by a complicated system of groups and filaments very similar to the other massive systems in the HRS supercluster (e.g., A3128/A3125, \citealp{rgc02}). At least three of the detected structures (2, 5, \& 6) appear to reside on the edge of the ICM. These relatively compact structures in the plane of sky have broad velocity distributions, well over the typical velocity dispersion of compact groups (400~\kms, \citealp{sr91}), and are possibly tidally disrupted groups impacted by the cluster's massive gravitational field. Alternatively, these structures might represent a series of merging groups during hierarchical structure build-up. Whichever the scenario for the main merger, these groups and filaments suggest A3266 is undergoing a series of complex dynamical interactions. 

\section{Summary}\label{sec:summary}

In summary, we present a detailed spectroscopic structure analysis of Abell 3266. We conducted spectroscopic observations over a $2\degr$-field surrounding A3266. We measured a total of 880 redshifts including 704 of which were new measurements. Combining these with the literature gave a total sample of 1303 of which 790 were in the cluster sample. This is the most complete spectroscopic sample of A3266 to date and allows detection of a range of substructures associated with the cluster. In particular, we detect 6 new structures surrounding the cluster core, and find the core to be split into two or three components with the main merger at about $120\degr$ clockwise from North. This is consistent with the detailed X-ray analysis of \citet{fhm06}. However, it is clear from this analysis that A3266 is an extremely complex system which cannot be explained by a simple merger scenario. An array of cluster merger simulations will be required to determine the most likely merger scenario and it is clear this will likely require multiple mergers to adequately fit the data. The complex merger history and amount of subclusters suggest a rich environment which would include shocks and turbulence. Such conditions along with a high cluster mass have been shown to be required for the generation of radio relics and halos \citep{cbn12,sjp16} and this will be explored in the second paper in this series.

\section*{Acknowledgements}

MJ-H is supported in this work by a Marsden Grant administered by the Royal Society of New Zealand. MC thanks to Ron Hola, Ian Lewis, Terry Bridges and Russell Cannon for assistance with the 2dF observations, and to Jess O'Brien for re-reducing the 1997 data. Steve Maddox kindly provided the astrometry and photometry from the APM galaxy survey.

The Digitized Sky Surveys were produced at the Space Telescope Science Institute under U.S. Government grant NAG W-2166. The images of these surveys are based on photographic data obtained using the Oschin Schmidt Telescope on Palomar Mountain and the UK Schmidt Telescope. The plates were processed into the present compressed digital form with the permission of these institutions.

This research has made use of the NASA/IPAC Extragalactic Database (NED), which is operated by the Jet Propulsion Laboratory, California Institute of Technology, under contract with the National Aeronautics and Space Administration.

This research has made use of NASA's Astrophysics Data System.




\bibliographystyle{mnras}
\bibliography{A3226Bib}{}

\bsp	
\label{lastpage}
\end{document}